\documentclass[final,11pt,authoryear]{elsarticle}

\usepackage{amssymb}
\usepackage{enumitem}
\usepackage{tabularx}
\usepackage{verbatim}

\journal{Journal of Choice Modelling}
\RequirePackage[normalem]{ulem} 
\RequirePackage{color}\definecolor{RED}{rgb}{1,0,0}\definecolor{BLUE}{rgb}{0,0,1} 

\begin{document}
\begin{frontmatter}

\title{Understanding the decision-making process of choice modellers}

\author[inst1]{Gabriel Nova\corref{cor1}}

\affiliation[inst1]{
            organization={Department of Engineering Systems and Services, TU Delft}, 
            country={Netherlands}}

\author[inst1]{Sander van Cranenburgh}

\author[inst1,inst2]{Stephane Hess}

\affiliation[inst2]{
            organization={Choice Modelling Centre - Institute for Transport Studies, University of Leeds}, 
            country={England}}

\begin{abstract}
Discrete Choice Modelling (DCM) serves as a robust framework for modelling human choice behaviour across various disciplines. Building a choice model is a semi-structured research process that involves a combination of a priori assumptions, behavioural theories, and statistical methods. This complex set of decisions, coupled with diverse workflows, can lead to substantial variability in model outcomes. To investigate these dynamics, we introduce the Serious Choice Modelling Game (DCM-SG) as a methodological approach that simulates the modelling process and records modellers’ decisions in real time. Participants were asked to develop choice models to estimate Willingness-to-Pay values to inform policymakers on strategies for reducing noise pollution using a stated preference dataset. Their actions were tracked across research phases, which allows us to analyse workflows patterns and modelling strategies. A total of 40 participants completes the game, 77\% of whom had more than five years of experience in DCM. While our findings reveal a strong preference for using data visualisation tools and frequent use of simpler models such as Multinomial Logit, there were also attempts to specify more complex models. These findings suggest that, in time-constrained or resource limited settings, modellers may underexplore important modelling factors such as covariates, non-linear transformations, and complex utility specifications.  Moreover, we observe that participants who engaged in more comprehensive data exploration and iterative model comparison tended to achieve better model fit and parsimony. These results demonstrate how sequential data obtained from the DCM-SG can be used to reveal variations in modelling practices and offer a starting point for understanding the decision-making within these processes.\end{abstract}

\end{frontmatter}
\clearpage

\section{Introduction}
\label{sec:Introduction}
Discrete Choice Theory (DCT) is a theoretical framework used across various scientific disciplines to study human choice behaviour. These fields include but are not limited to, transport, health, and environmental economics \citep{louviere2000stated, hess2024handbook, mariel2021environmental, haghani2021hypothetical}. This theory informs analysts on how to specify Discrete Choice Models (DCMs) for estimating and predicting choices \citep{ben1985discrete}. On the one hand, by calibrating models on empirical choice data, choice modellers can estimate and infer preferences over alternatives and their attributes, which correspond to features or qualities that define them. On the other hand, by using the estimated parameters to simulate choice scenarios, they can predict future behaviour and responses to changes in policy or market conditions. This allows analysts not only to study the decision-making process and the factors that influence individual decisions, but also to analyse choice behaviour in different contexts, forecast demand, and evaluate policy changes (Ben-Akiva \& Bierlaire, 2003).\\

\noindent Discrete choice modelling brings together individuals with diverse backgrounds and expertise to understand and forecast choice behaviour through a series of research steps. Whatever the purpose, choice modellers typically engage in workflows that involve formulating a research question, experimental design, data collection, data exploration, descriptive analysis, model specification, outcome interpretation, and reporting \citep{ben1985discrete, hensher2015applied, mariel2021environmental}. Throughout each research phase, these professionals balance various formal behavioural theories with statistical methods, experimental applications, their own knowledge and professional judgements to develop models that represent the choices in the data under study \citep{paz2019specification}. Although these phases generally follow a chronological order, they are often carried out in a semi-structured manner. This flexibility enables modellers to determine how and when to incorporate subjective knowledge acquired during the process, ultimately guiding the selection of a model after evaluating multiple specifications \citep{rodrigues2020bayesian, van2022choice}. This inherent flexibility, combined with different workflows and subjective decision-making, can lead to model specifications that do not truly capture the data generation process. Modellers may interpret data differently, emphasise different model aspects on the functional forms, or even select model families that do not reflect the preferences in observed choices, which in turn leads to considerable variability in modelling results and conclusions. For this reason, choice modelling is often considered an art, requiring decision-making with a high degree of freedom that allows modellers to use their expertise to make decisions within their research.\\

\noindent Hitherto, studies have focused primarily on the model specification phase
\citep{daly2012calculating, mcfadden1974measurement, mcfadden2000mixed, walker2007latent}. This phase involves a trial-and-error process where choice modellers determine both the model structure and the parameters to be considered in the model \citep{paz2019specification}. For instance, analysts must decide on the choice of model family, the  inclusion of linear or non-linear transformations to variables, the incorporation of observed heterogeneity, the distributions of the random coefficients and their correlations, among other considerations 
\citep{train2009discrete, beeramoole2023extensive,mariel2021environmental}. This iterative and time-consuming process continues until modellers have estimated each specification, and obtained goodness-of-fit indicators and validation metrics to assess model fit, parameter consistency, and alignment with existing literature \citep{parady2021overreliance}. As modellers may use different cost functions to balance these goodness-of-fit metrics, this process may end with various specifications that they consider acceptable to address their research question. \\

\noindent The current choice modelling landscape reveals a knowledge gap in understanding modellers’ decision-making processes. Despite recent developments have introduced algorithms designed to assist in the model specification (typically using goodness-of-fit indicators as objective) \citep{paz2019specification,ortelli2021assisted,beeramoole2023extensive}, these algorithms do not take into account relevant aspects of the modelling process and only partially replace some model specification decisions.  They do not consider the modeller's decision-making during the descriptive analysis phase, the trade-offs made during model specification to constrain the search space, nor the model selection at the end of this process. This inherent flexibility of choice modellers' workflows can lead to diverse results, interpretations, and conclusions even when working with the same choice dataset. Similar concerns have been observed in psychology, where concepts such as ‘researcher degrees of freedom’ 
\citep{simmons2011false} and the ‘garden of forking paths’ \citep{gelman2013garden} emphasise how flexibility in data collection and multiple potential tests based on the same dataset, along with the pursuit of meaningful parameters, can increase the risk of false positives. Furthermore, crowd-science experiments have demonstrated significant variability in research processes, highlighting a lack of consensus in decision-making, and divergent outcomes when different researchers analyse the same data 
\citep{botvinik2020variability, wicherts2016degrees, silberzahn2018many}
. While this degree of freedom promotes exploration and methodological innovation, it also carries a risk of poor decisions and undesirable outcomes. A better understanding of these processes not only encourages debate about best practices, but also paves the way for improved practices within the modelling community.\\

\noindent This study aims to shed light on the choice modelling research process by introducing a novel methodological approach that combines serious game design, empirical data collection, and sequential pattern mining analysis. Specifically,  it presents the first serious game designed to simulate the modelling process and capture real-time decision-making by choice modellers within a controlled research environment. The game covers the full process from data exploration and model specification to outcome interpretation and results reporting. It was applied among participants from two internal conferences and additional online recruits, with their actions tracked throughout. Rather than evaluating participants against a predefined modelling strategy, our aim is to reveal how modellers navigate the modelling process, respond to feedback, and iterate on their modelling assumptions. These data then provide a unique opportunity to explore modelling workflows, reveal the degrees of freedom available to choice modellers, and analyse how their use of in-game tools and workflows influence reported results. In doing so, this study not only contributes a methodological instrument for analysing decision-making in choice modelling, but also demonstrates how sequential data derived from the game can reveal differences in modelling practices and serve as a starting point for understanding these workflows.\\

\noindent The remainder of the paper is structured as follows. Section \ref{sec:Related-Work} discusses related work, focusing on the progression from data analysis to modelling results. It also introduces a conceptual framework, which outlines the modelling phases that guide our serious game design. Section \ref{sec:Method} describes the methodology, covering the serious game  design, the in-game tools available at each choice modelling phase, and the stated preference choice dataset used. While Section \ref{sec:SG-data} details the gameplay session, Section 5 presents the results and discussion of the modellers' workflows. Finally, Section \ref{sec:conclusions} presents the main conclusions.

\section{Theoretical framework : Beyond data analysis to modelling}
\label{sec:Related-Work}
Discrete choice modelling (DCM) enables choice modellers to uncover preferences and forecast choices across different disciplines. Unlike conventional data analysis approaches, DCM involves not only statistical techniques, but also requires analysts to build, evaluate, and validate models that capture the underlying trade-offs of individuals in making choices\citep{ben1985discrete}. Furthermore, while some modellers focus on developing and testing theoretical frameworks, others apply these models to real-world datasets to gain insights into specific choice contexts. Their tasks include collecting and analysing choice data, developing and validating models, interpreting results and calculating econometric metrics to better represent decision-making and understand factors that influence choice behaviour. Given the complexities involved in the modelling process, this section presents a general framework that outlines the modellers' degrees of freedom and main research phases for developing choice models, based on well-established literature (e.g., \cite{ben1985discrete, train2009discrete, louviere2000stated, mariel2021environmental, hess2024handbook}; among others).

\begin{figure}[h!]
    \centering
    \includegraphics[width=\textwidth]{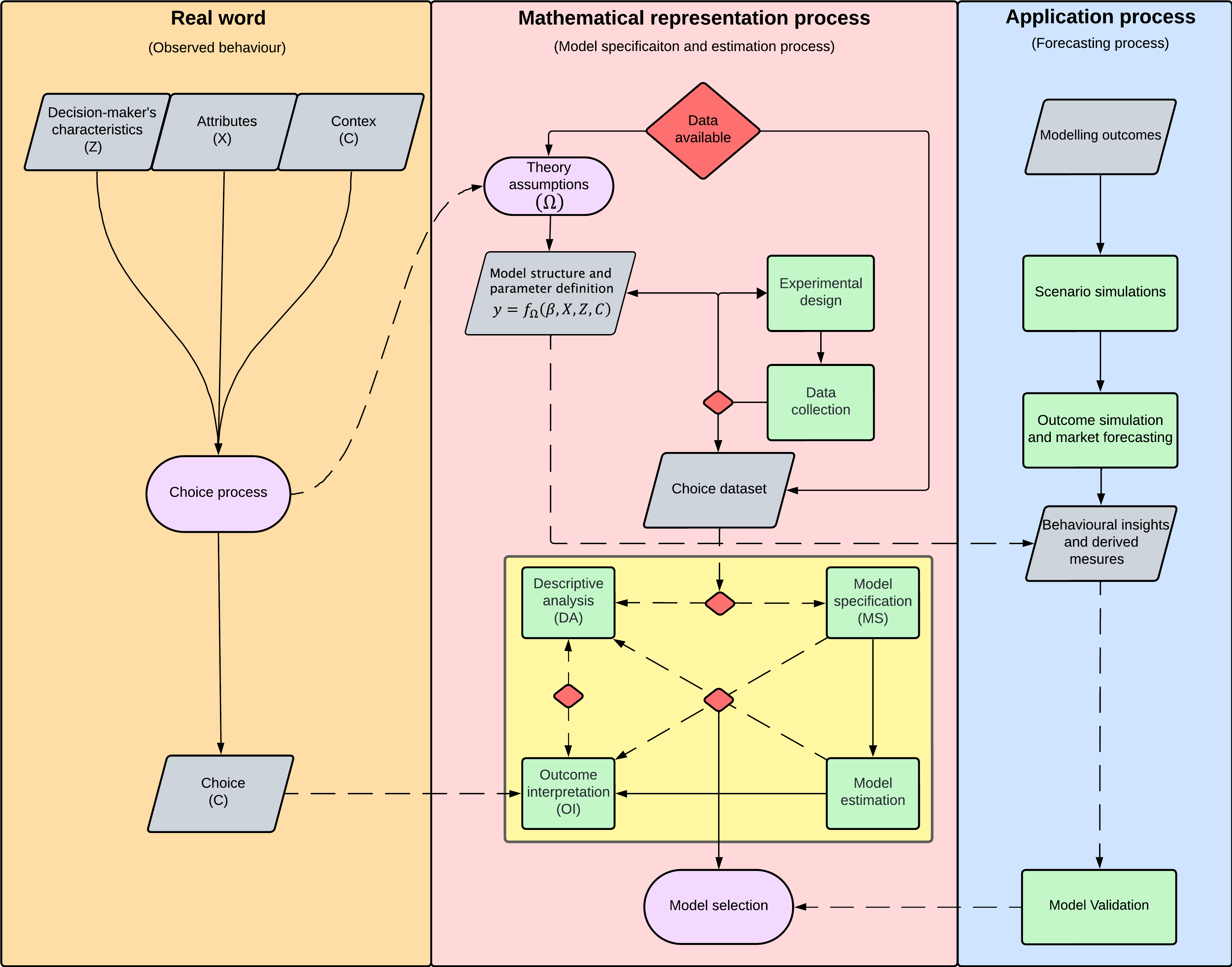}
    \caption{Conceptual overview of the choice modelling research process}
    \label{fig:DCM_overview}
\end{figure}
\noindent Figure \ref{fig:DCM_overview} presents a theoretical framework for discrete choice modelling, illustrating the conceptual flow from observed choice behaviour to formal model representation and application cases. The framework is organised into three interconnected processes, each demarcated by vertical sections: the real-world choice process (left), the mathematical representation process (centre), and the application process (right). In addition, processes are represented by rectangles; data, inputs, or outputs are depicted by parallelograms; internal processes performed by decision-makers or choice modellers are shown as ovals; and modellers' decisions are indicated by diamonds. Solid arrows convey sequential progression, while dashed arrows highlight the influence between components.\\

\noindent In the real world (left), decision-makers' choice process is represented using the canonical representation, where decision-makers with specific characteristics (Z) face a choice situation defined by alternatives and their attribute levels (X) within a given hypothetical context (C). These inputs guide the decision-maker's choice process, which ultimately results in an observable choice outcome (y) that modellers aim to understand and forecast. \\

\noindent The mathematical representation process captures the entire modelling process workflow, including theoretical formulation, data handling and exploration, model estimation, and selection. To model real-world choices, modellers must begin with a choice dataset to address their initial research question, which is guided by data availability. If a suitable dataset already exists, choice modellers may proceed directly to data handling and exploration. However, in the absence of a dataset, several decisions about experimental design and data collection methodologies must be made, guided by modellers' theoretical assumptions \citep{ben1985discrete}.\\

\noindent Experimental design and data collection require modellers to make several decisions, often influenced by their theoretical assumptions and subjective judgements that shape the resulting dataset. 
Although discrete choice models can be estimated using revealed preferences (RP), stated preferences (SP), or both, the focus of this study is exclusively on SP data. RP involves observing actual choices through methods such as self-reported data, passive data collection, activity diaries, or even psychophysiological measures
\citep{bierlaire2008route,arriagada2022unveiling,hancock2022secret, barria2023relating, xu2018real}. For instance, in transport modelling, RP data collection involves determining the basic unit of study, the spatial range, and essential information such as origin, destination, purpose, start time, end time, payment method, transport mode, and sociodemographic variables \citep{axhausen2024self}. While RP data reflects real behaviour, its design is inherently limited to existing and observed alternatives, making it difficult to evaluate preferences for hypothetical or future scenarios.\\

\noindent In contrast, SP experiments allow modellers to design hypothetical choice tasks to capture preference and choice behaviour under controlled conditions. Designing such experiments introduces a series of sequential decisions that not only influence the ability to identify primary attribute effects but also influence the statistical power to find significant relationships between variables, particularly in studies with typical sample sizes \citep{Rose2009}. These decisions include defining the number of choice tasks, the alternatives, the attributes, and their levels \citep{hensher2004identifying, caussade2005assessing, Rose2009, meyerhoff2015influence, meissner2016eye, mariel2021environmental}.  Moreover, the RP data can be used to inform the SP design by suggesting attribute range values or common trade-offs that lead to a more realistic choice task. The survey instrument is then constructed and potentially refined through pilot studies or focus groups.\\

\noindent The resulting dataset is used across several interconnected modelling phases: descriptive analysis (DA), model specification (MS), model estimation (ME), and outcome interpretation (OI). These phases are dependent through a feedback loop that reflects the trial-and-error nature of refining model assumptions and model specifications until a suitable model or multiple models are selected. It is widely known that most modellers begin this process with exploratory and descriptive analysis, which is essential for understanding the structure and composition of the data, as well as for preprocessing it for subsequent modelling tasks. This phase includes statistical analysis, graphical representation, handling missing values, and dummy coding or scaling to prepare the data for model estimation. Beyond these analyses, exploratory analysis plays an important role in revealing relationships and correlations between attributes, covariates, and outcomes. These insights guide the formulation of initial modelling hypotheses, providing a sense of which variables to include, how they may interact, and what behavioural patterns are being modelled. Although some modellers might skip this phase to save time,  doing so they risk missing valuable information that can significantly improve model specification.\\

\noindent Model specification is a pivotal phase where modellers translate theoretical assumptions into a formal utility representation, deciding both the functional form and its behavioural mechanisms to be captured. Modellers must choose a model family depending on the distribution of error components and heterogeneity across decision-makers (e.g., Multinomial Logit, Generalised Extreme Value, Latent Class, Mixed Logit). Then, modellers must determine the model specification itself, including which variables to include, how to include them (e.g., linear, logarithmic, piecewise), and whether to incorporate interaction with covariates (e.g., Gender, Age, Household composition) \citep{ortelli2021assisted,rodrigues2020bayesian}. These degree of freedom are not only essential for capturing the complexity of real-world choices  \citep{beeramoole2023extensive}, but also influence both model interpretability and performance \citep{parady2021overreliance, van2022choice}. This is an iterative process that involves estimating various model specifications, validating them internally and externally, and analysing and interpreting modelling outcomes by comparing model fit, parameter consistency, and alignment with existing literature (for more details, see \cite{parady2021overreliance}).\\

\noindent In the final phase, the selected model(s) are applied to generate behavioural insights, forecast demand, and evaluate policies in different contexts. This involves using the estimated parameters to simulate choice scenarios, derive measures such as willingness-to-pay, elasticities, and market share, among others, and predict responses to changes in policy or market conditions. The application process allows choice modellers to derive practical implications from their models, providing valuable insights for policymakers.\\

\noindent To summarise, the choice modelling framework provides a structured approach to understanding and forecasting individual choice behaviour by capturing the choice process through several model specifications. Beyond data collection and analysis, the numerous workflows and decisions in model specification, estimation, outcome interpretation, and model comparison can create an opaque environment for both understanding the actual data generation process and the factors that influence decision-makers in the choice situation studied by choice modellers. This has significant implications not only for the field itself but also for the reliability of the modelling outcomes, which policymakers rely on for policy formulation and analysis. Enhancing transparency and addressing these degrees of freedom can strengthen the faith that policymakers place in the results provided by choice modellers.

\section{Method}
\label{sec:Method}

\subsection{Serious games for choice modelling}\label{sec:3.1}
Serious Games (SG) game tools with a primary purpose distinct from entertainment, particularly for addressing real-world problem that provide training, learning or behavioural change for participants. \citep{michael2005serious}. They are characterised by explicit rules and defined goals, SGs are intentionally designed to be applied to relevant issues where players can experiment in a safe environment with different in-game tools and see the consequences of their decisions \citep{corti2006games, dorner2016introduction, squire2006content} . Although initial applications aimed to improve the user’s decision-making skills in a diverse simulated environment, such as educational \citep{rawitsch1978oregon} and military combat performance \citep{krulak1997military}, later efforts have concentrated on tracing players’ actions and behaviours within the game environments \citep{medler2011analytics}. \\

\noindent SGs have been gaining popularity in behavioural evaluation due to their potential to capture data in real-time and allow analysts to use them to perform decision-making analysis in different problems, such as policy formulation, resource allocation, or scenario analysis \citep{donaldson2002understanding, olejniczak2020advancing, van2021experimental}. Although data collection can be conducted during different phases, such as pre-, in- and post-game, which offers a variety of methods and types of data \citep{mayer2014research}, most early serious games do not focus on collecting in-situ information on player behaviour \citep{smith2015meta}. Surveys, questionnaires, and self-reports were the most commonly used methods due to their simplicity in terms of data collection. However, these fail to provide detailed insights into decision-making processes, as participants tend to modify their behaviour during the game to align with the researchers' objectives \citep{podsakoff2003common} and do not capture participants' actual thinking processes
\citep{lazar2017research}. Therefore, with technological progress, efforts have been made to collect behavioural data during gameplay in simulated environments with controlled variables that can be used to demonstrate relationships between different players' knowledge, degrees of game experience, or backgrounds, with their performance in different tasks \citep{chung2014toward, snow2014entropy}.\\

\noindent Integrating SG into choice modelling offers a novel method for understanding the decision-making process, viewed as a series of interconnected and complex research phases. Thus, the Discrete Choice Modelling Serious Game (DCM-SG) was designed based on the conceptual model outlined in Section \ref{sec:Related-Work}, assuming the availability of a collected SP choice dataset. Based on this, the game allows users to move through the phases of descriptive analysis, model specification, estimation, and outcome interpretation, with the objective of deriving measures such as willingness-to-pay estimates. Indeed, being able to capture choice modellers’ decisions in real-time data facilitates an analysis of actual behaviour throughout the performed workflows. Furthermore, the data collected enables an analysis of the relationships between participants' experience, the field of expertise, and knowledge, and how these factors influence their decision-making processes.

\subsection{Discrete Choice Modelling Serious Game design}\label{sec:3.2}
To reveal how choice modellers make decision throughout choice modelling process, we designed and developed the Discrete Choice Modelling Serious Game (DCM-SG) based on the framework proposed by \citet{meijer2009organisation}. The game is intended for students, researchers, practitioners, and analysts with at least a basic understanding of choice modelling. No programming skills are required. All modelling actions are performed through a user-friendly interface, where predefined triggers guide participants through research phases such as data exploration, model specification, and outcome interpretation.\\

\begin{figure}[h!]
    \centering
    \includegraphics[width=\textwidth]{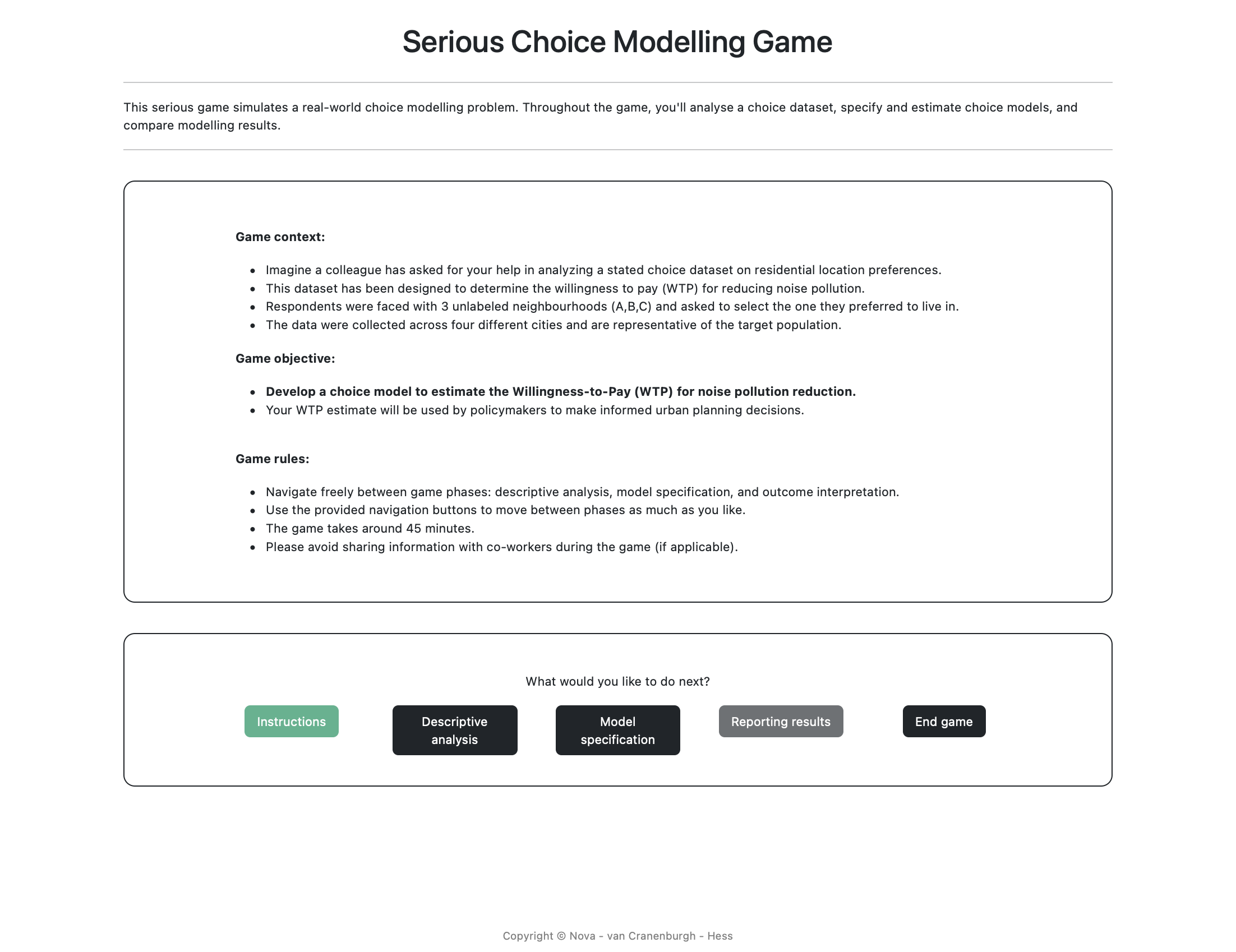}
    \caption{Screenshot of the DCM-SG instruction page}
    \label{fig:Instructions}
\end{figure}

\noindent During the game session, participants are assigned the role of choice modellers and the game is played with a reference system. Our reference system is confined to a real-world problem that involves developing choice models using a stated preference choice dataset, where participants with different backgrounds may apply different approaches to solve this problem. Participants are presented with a well-defined context as “\textit{Imagine a colleague has asked for your help in analysing a stated choice dataset on residential location preferences. This dataset has been designed to determine the willingness to pay (WTP) for reducing noise pollution. Respondents were faced with 3 unlabeled neighbourhoods (A, B, C) and asked to select the one they preferred to live in. The data were collected across four different cities and are representative of the target population}”. Thus, they are required to apply their knowledge, skills, and non-explicit competencies to generate and deliver modelling outcomes to inform policymakers, while their actions and responses are tracked. Specifically, we stated the game objective as “Develop a choice model to estimate the Willingness-to-Pay (WTP) for noise pollution reduction. Your WTP estimate will be used by policymakers to make informed urban planning decisions”, as shown on the instruction page (Figure \ref{fig:Instructions}). Therefore, this problem-solving context situates our method within the testing-and-retention quadrant, as described by \cite{olejniczak2020advancing}.\\

\noindent To ensure that the game’s design was appealing and as realistic as possible, we also specified clear rules and constraints throughout the game experience. These were explained at the beginning of the game sessions, and also appear within the game instructions themselves. On the one hand, the rules state that participants can perform any allowable actions within the game, whilst avoiding the sharing of information or results with other participants during the 45-minute game session. On the other hand, some constraints were introduced to limit the degrees of freedom that exist in the model specification phase.

\subsection{DCM-SG tools}\label{sec:3.3}
To facilitate participants' workflows in solving this problem, the game simulates four main phases of choice modelling research, derived from the conceptual framework: Descriptive analysis (DA), Model specification (MS), Outcome interpretation (OI), and Reporting (R). Participants can iterate through each phase as many times as they deem necessary before moving on to the next or previous phase, or even returning to phases already played, thus mimicking the intrinsic trial-and-error phenomenon of modelling process. The iterative approach, which aligns with the conceptual framework depicted in Figure \ref{fig:DCM_overview}, enables participants to continuously refine their specifications and analyses until they report their findings.\\

\begin{figure}[h!]
    \centering
    \includegraphics[width=\textwidth]{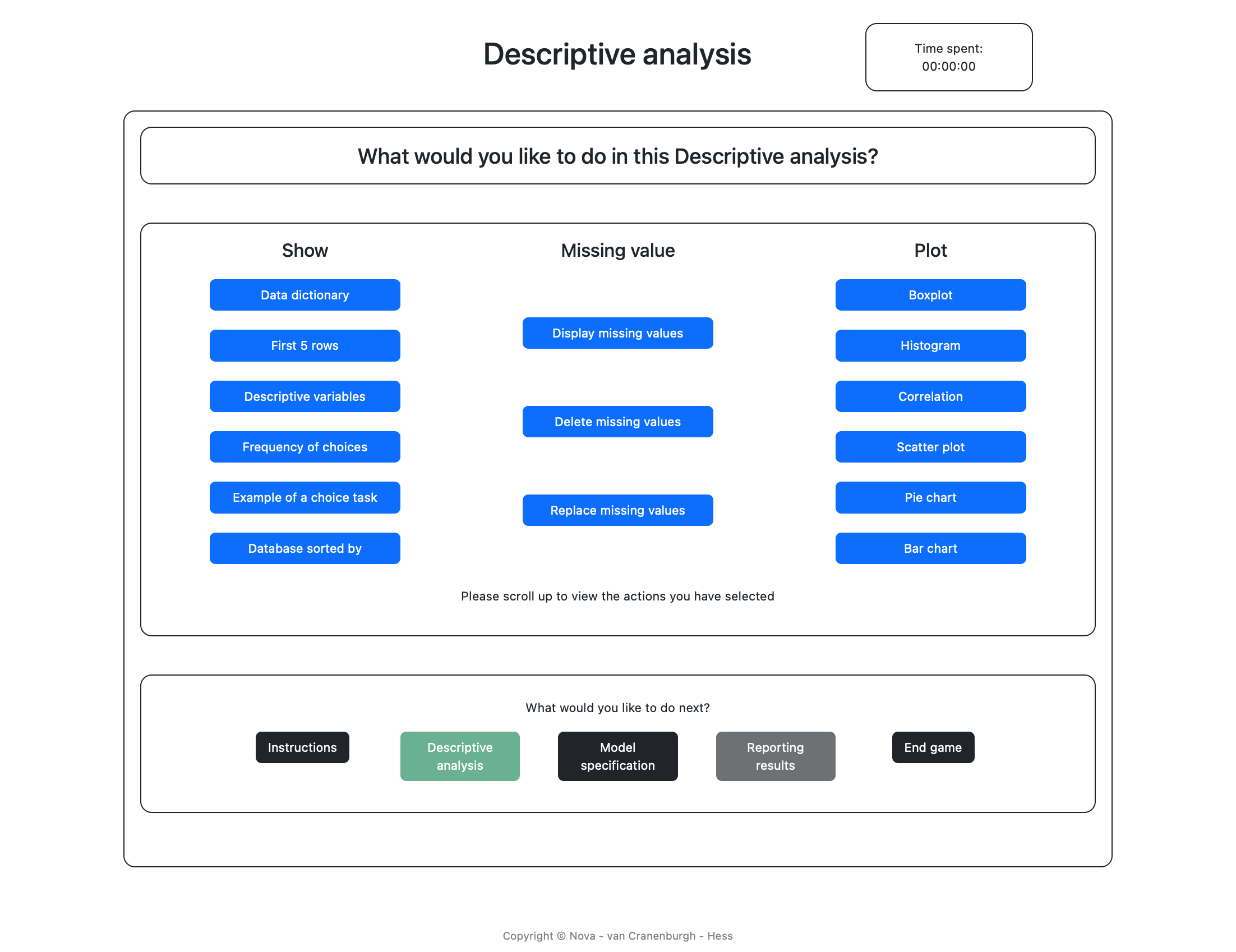}
    \caption{Screenshot of the DCM-SG descriptive analysis page}
    \label{fig:descriptive_analysis}
\end{figure}

\noindent \textbf{Descriptive analysis}: During this phase, we incorporated in-game tools based on general practices in choice modelling and recommendations from the literature \citep{tukey1977exploratory, paez2022discrete}. This allows modellers to perform a range of exploratory data analysis tasks on the raw choice dataset, which enables them to better understand the data distribution prior to the model specification phase. Specifically, participants  can view a data dictionary to understand the context and meaning of each variable; inspect the first five rows of the dataset;  consult descriptive statistics (e.g., mean, median, max, min, standard deviation) to obtain central tendencies. Participants can also examine the frequency of choices (distribution of choices made by decision-makers) and see an example of a choice task. Additionally, they can sort the dataset by any variables to see the completed dataset. For handling missing data, participants can display missing values, delete missing values, or replace them (using the mean, mode or median). Finally, the visualisation tools available include box plots, histograms, correlation matrices, scatter plots, pie charts, and bar charts, which allow users to plot any variable and explore the structure of the dataset. While we acknowledge that the list of actions is not exhaustive, these tools were selected to reflect techniques commonly used in the field and to maintain consistency among participants. Thus, our serious game ensures comparability across workflows and highlights how participants engage with the data under similar conditions. An overview of this phase is shown in Figure \ref{fig:descriptive_analysis}.\\

\begin{figure}[h!]
    \centering
    \includegraphics[width=\textwidth]{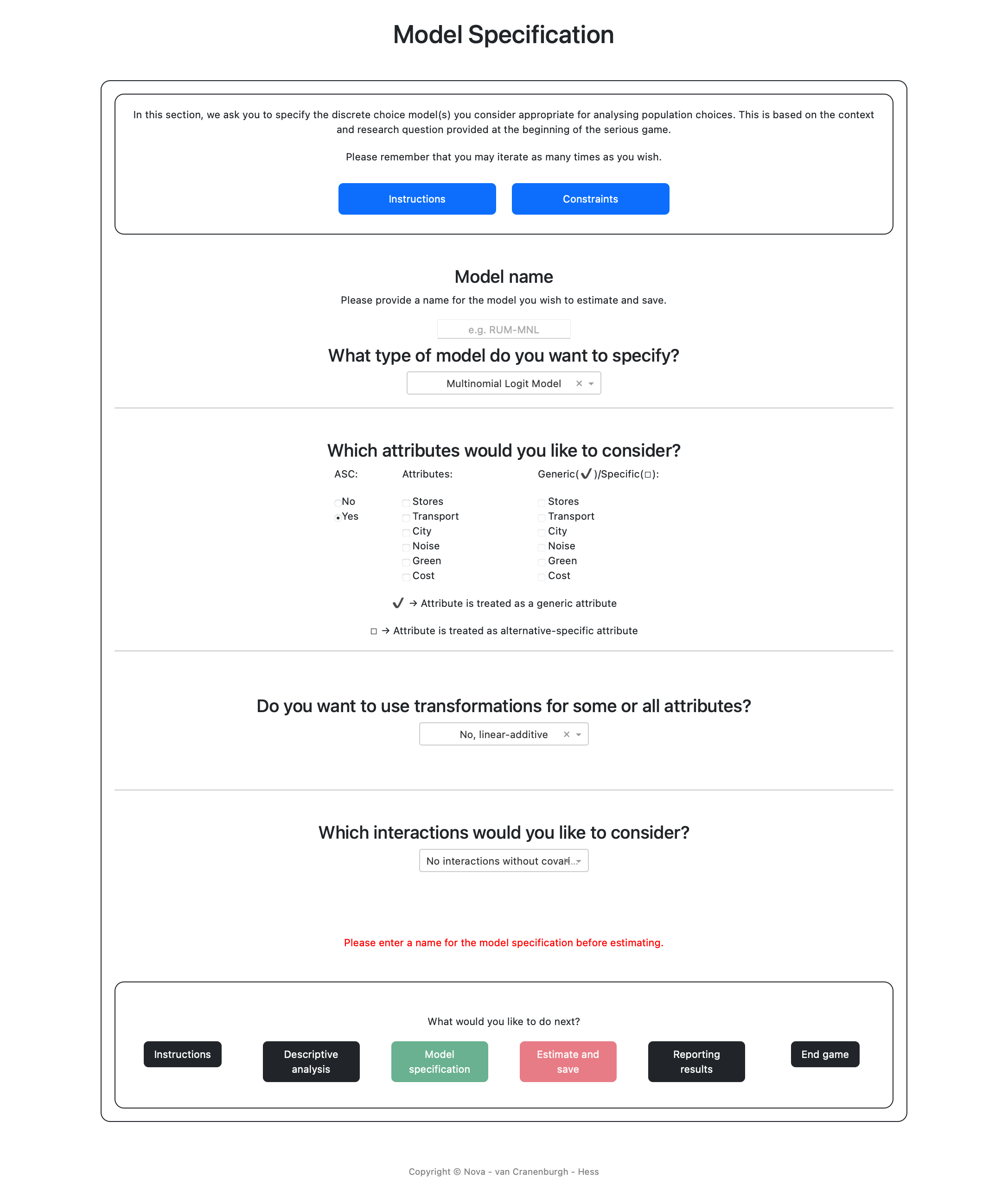}
    \caption{Screenshot of the DCM-SG model specification page}
    \label{fig:model_specification}
\end{figure}

\noindent \textbf{Model specification}: During this phase, researchers are given the flexibility to specify models from diverse families that are foundational and widely used in discrete choice modelling, as shown on the respective page (Figure \ref{fig:model_specification}). The game includes the Multinomial Logit (MNL) model 
\citep{mcfadden1974measurement}, which is considered the workhorse and the benchmark due to its simplicity, interpretability, and capacity for capturing population taste variation in observed variables and relative substitution between alternatives 
\citep{train2009discrete}. To specify an MNL model, modellers need to make various decisions, such as incorporating alternative-specific constants, including attributes, selecting between generic or alternative-specific coefficients, and considering interactions with sociodemographic variables. They can also apply non-linear transformations, such as logarithmic, and Box-Cox functions, to account for non-linearities.\\

\noindent To capture unobserved heterogeneity, the DCM-SG also allows the specification of the Mixed Multinomial  Logit (MMNL) model 
\citep{mcfadden2000mixed}, which is commonly used to address random taste variation across individuals and correlation in unobserved factors, and does not satisfy Independence from Irrelevant Alternatives property, making it powerful to model a range of behavioural specifications 
\citep{train2009discrete}. Within this model family, modellers can decide which parameters follow a random distribution (normal or lognormal) while maintaining the remaining ones as population-level parameters. Additionally, they can account for observed heterogeneity by interacting attributes with sociodemographic variables. Due to the exponential number of specifications a modeller might consider, some limitations are introduced for MMNL models:

\begin{enumerate}[noitemsep]
    \item Alternative-specific constants are considered for all models.
    \item All attributes are included in the utility function and treated generically across alternatives.
    \item The number of draws is fixed and constant across all models, and this aspect is not analysed.
    \item A maximum of two random parameters can be included in the utility specification.
    \item Interaction between a random attribute and a sociodemographic variable is not allowed.
\end{enumerate}

\noindent The serious game also considers Latent Class Models \citep{walker2007latent}, which enable the segmentation of decision-makers into discrete classes by assuming that each class have distinct preferences. This model is included since it is among the most extensively used in the choice modelling literature
\citep{hensher2015applied}. In these model specifications, modellers not only have to define the number of latent classes, but also have to determine which covariates to include in the membership class to calculate the probability of belonging to each class. Similar to the MMNL case, some limitations are introduced:

\begin{enumerate}[noitemsep]
    \item Alternative-specific constants are considered.
    \item All attributes are included in the utility function and treated generically across alternatives.
    \item Models can be specified with up to three latent classes.
    \item Covariates are dummy coded and may be included in the class allocation model.
\end{enumerate}

\noindent Furthermore, we decided not to include machine learning (ML) models. While we acknowledge the growing interest in models such as Random Forests, Support Vector Machines, Gradient Boosting Decision Trees, and Artificial Neural Networks for analysing choice preferences, their adoption is still relatively limited \citep{hagenauer2017comparative, wang2017travel,wang2020deep,martin2023prediction}.  We believe that including ML  models would introduce unnecesary complexity rather than enhance realism of the SG. \\

\noindent Finally, regarding the estimation of the models. MNL models are estimated on the fly upon user request. To do so, the DCM-SG integrates Biogeme methods and classes, allowing users to dynamically request and estimate any specification of this model family according to their specific needs. Due to the longer estimation times required for MMNL and LC models, we have pre-estimated using the Delft Blue
\citep{DelftBlue2024} and local machines, employing both Apollo \citep{hess2019apollo} and and Biogeme \citep{bierlaire2003biogeme}. The results generated by these software packages were stored in a database to serve as a repository for the outcome interpretation phase. In total, we have pre-estimated 78,604 MMNL and 8,832 LC models. Lastly, we calculated the standard errors of the Willingness-to-Pay estimates using the Delta method \citep{daly2012calculating}. The Willingness-to-Pay values and standard errors are available to the participants in the outcome interpretation phase.\\

\begin{figure}[h!]
    \centering
    \includegraphics[width=\textwidth]{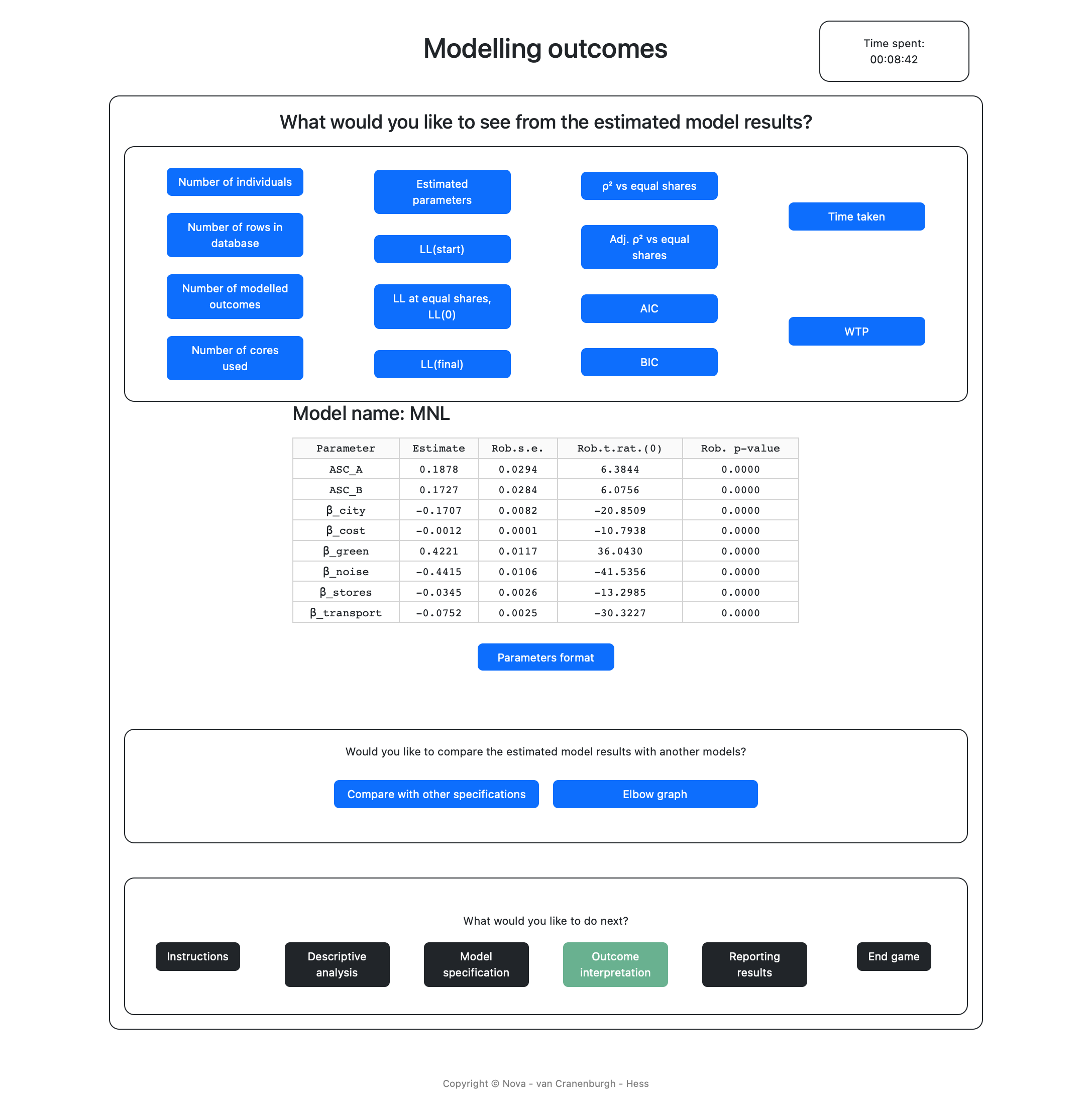}
    \caption{Screenshot of the DCM-SG report page}
    \label{fig:Outcome_interpretation}
\end{figure}

\noindent \textbf{Outcome interpretation}: During this phase, we display the common modelling result table for the estimated model, which contains parameter names along with their estimated values and standard error, t-test, and p-values, as shown on the respective page (Figure \ref{fig:Outcome_interpretation}). These results represent the most elementary outputs following model estimation, which provide the information to assess the significance of parameters and their alignment with theoretical expectations. Modellers thus typically consider these results before examining additional goodness-of-fit metrics. Then, participants can decide on a range of metrics for the estimated model, and decide to analyse them, such as the number of parameters estimated in the model, the size of the sample used in the estimation process, the null log-likelihood, the initial log-likelihood, the final log-likelihood, the likelihood ratio test against the null model, the rho-squared against the null model, the adjusted rho-squared against the null model, the Akaike Information Criterion, Bayesian Information Criterion, final gradient norm, time taken for model estimation, and Willingness-to-Pay for attribute estimates. Finally, they can compare models in two manners: by making direct comparisons between models in terms of parameters and goodness-of-fit indicators, or by displaying elbow graphs to visualise latent class model metrics.\\

\begin{figure}[h!]
    \centering
    \includegraphics[width=\textwidth]{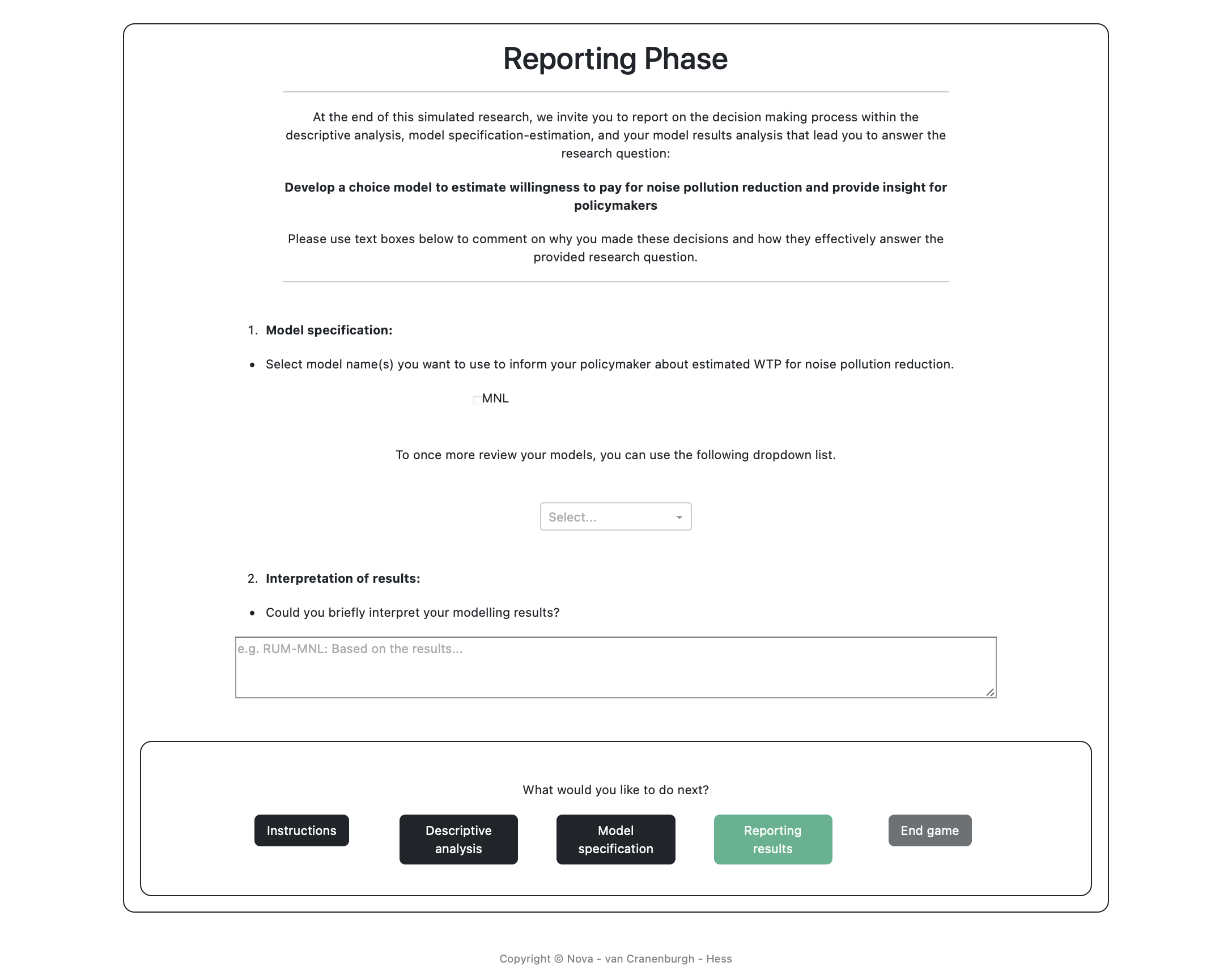}
    \caption{Screenshot of the DCM-SG report page}
    \label{fig:report}
\end{figure}

\noindent \textbf{Reporting phase}: At the end of the simulated research, choice modellers are asked to report their findings to policymakers, as shown on the respective page (Figure \ref{fig:report}). To facilitate this, they can review the estimated models along with summaries of their results, in order to select the most appropriate ones. In addition, participants are required to submit a short written report (several sentences), in which they detail the main findings and interpret the modelling results to address the objective set at the beginning of the game.

\subsection{Stated preference choice dataset}\label{sec:3.4}

The above discussion presents a general overview of the DCM-SG, independently of the dataset used. For our specific applications, and in order to represent a research scenario that imitates what practitioners deal with in their real-world work, we use a modified raw stated preference dataset collected by \cite{liebe2023maximizing}, which aimed to analyse residential location choice. The dataset consists of three unlabelled alternatives (A, B and C), each defined by six attributes (distance to the grocery store, distance to transportation, distance to city centre, street traffic noise, green areas in the residential area, and monthly housing cost variation), This dataset consists of 2,430 individuals, each facing four choice tasks, resulting in 9,720 observations. This dataset also includes sociodemographic variables, such as age, gender, home ownership, car ownership, residence, and employment status. A summary of the data is shown in Table \ref{T:Data}.

\begin{table}[h!]
\centering
\begin{tabularx}{\textwidth}{lXl}
\hline
\textbf{Variable} & \textbf{Description} & \textbf{Type/Levels} \\ \hline
ID & This is the ID number of the respondent & Integer \\
Task ID & This is the ID number of the choice task & Integer \\
Stores & Distance to grocery store in walking time & 2, 5, 10, 15 mins. \\
Transport & Distance to public transport stop in walking time & 2, 5, 10, 15 mins. \\
City & Distance to city centre in kms & $<$1, 1 to 2, 3 to 4, $>$4 km \\
Noise & Street traffic noise & None, Little, Medium, High \\
Green & Green areas in residential area & None, Few, Some, Many \\
Cost & Monthly change in housing cost \emph{vs} current & -€150, -€50, €50, €150 \\
Choice & Indicates the choice & 1 = A, 2 = B, 3 = C \\
Age & Age in years & $<$ 30, 30 to 50, $\geq$ 50 \\
Woman & Indicates if respondent is a woman (1) or not (0) & Binary \\
Homeowner & Indicates if respondent is a homeowner (1) or not (0) & Binary \\
Carowner & Indicates if respondent is a car owner (1) or not (0) & Binary \\
Respcity & Indicates corresponding city & Categorical \\
Job & Indicatest if respondent is working (1) or not (0) & Binary \\ \hline
\end{tabularx}
\caption{Data dictionary}
\label{T:Data}
\end{table}

\section{Serious game data}\label{sec:SG-data} \subsection{Gameplay data}\label{sec:4.1}
During gameplay in the DCM-SG, two types of data are recorded to enable behavioural analysis of decision-making in the choice modellers' workflows. Firstly, our game stores each participant's identifier, timestamp, and any task performed over the course of the research phases collected in situ and in real-time (telemetry). For example, there are 15 descriptive analysis in-game tools  that include statistical analysis, missing value handling, graph creation, and database sorting. In the model specification phase, there are 34 interactive game devices (such as buttons, drop-down menus, and checklists) for specifying MNL, MMNL, and LC models. During the outcome interpretation, there are 18 options to compare modelling results. Finally, the model(s) selected and their main findings are stored. Table \ref{T:gameplay} displays variables and their description, which allow us to track participants' interactions with the game.\\

\begin{table}[h!]
\centering
\begin{tabularx}{\textwidth}{lXl}
\hline
\textbf{Variable} & \textbf{Description} \\ \hline
timestamp & Time at which participants performed any task \\
user\_id & Participant ID \\
task\_id & Task ID performed in DA or OI\\
model\_id & Model ID (identify task performed in model specification) \\
model & 1 for MNL, 2 for MMNL, 3 for LC \\
ASC & 1 to include alternative specific constants (0 otherwise) \\
att$_i$ & 1 to include attribute i (0 otherwise) \\
s$_i$   & 1 to indicate that attribute i is alternative-specific (0 otherwise) \\
t$_i$   & 1 for applying linear transformation to attribute i, 2 for box-log, 3 for logarithmic \\
int$_i$  & indicates whether attribute i interacts with a single sociodemographic variable (only once per attribute is allowed at a time): none (= 0), woman (= 1), age (= 2), location of residence (= 3), homeowner (= 4), carowner (= 5) \\
dist$_i$ & 0 to indicate that attribute i is fixed at the population level, 1 follows a normal distribution, 2 follows a lognormal distribution \\
n\_class & Indicates the number of classes \\
covariates$_j$ & 1 to indicate that covariate j is included in the membership class function for latent class (0 otherwise) \\
r\_models & Identify reported models by the participants \\
reporting & Discussion and findings of selected models \\ \hline
\end{tabularx}
\caption{Variable Descriptions and Types}
\label{T:gameplay}
\end{table}

\noindent Secondly, both qualitative and quantitative data are collected at the end of the game to characterise participants in terms of their background and expertise. Thus, we characterised each participant using the user\_id and collected personal information such as gender (participant’s gender); age\_dcm (years involved in choice modelling); main\_field (primary field the participant is working in); and expertise (self-assessment of their knowledge in choice modelling). We also ask whether the participant has been a teacher or a student in a choice modelling course; their most commonly used programming language or software; their predominant modelling approach used in their applications; and their h\_index
(as a proxy for scholarly impact). Finally, we recorded the initial and end timestamp of when participants began playing the game (init\_time) and submitted their report (end\_time).

\subsection{Participants}\label{sec:4.2}
The DCM-SG was administered in person to attendees at two conferences, the International Choice Modelling Conference and the International Conference on Travel Behaviour Research, and was also distributed online to researchers and practitioners known to work with or on Choice Modelling. A total of 40 participants were involved, 38 of them reported their models and completed their analyses. For the remaining participants, we inferred their reported models as those that were specified toward the end of their modelling processes and demonstrated high performance. Although the distribution of experience in choice modelling is varied, most participants have more than 5 years of experience (10\% had less than 1 year, 32\% had 1 to 5 years, 45\% had 5 to 10 years, and 13\% had more than 10 years). Moreover, 85\% of the participants had taken a choice modelling course, while 58\% had experience as a teacher, teaching assistant, or lecturer. These values indicate that most participants were familiar with more than just the basics of choice modelling. Many could be considered experts, combining formal training, teaching experience, and several years of involvement in the modelling process. They mostly focused on transportation (70\%), while others worked in economics (10\%), urban planning (8\%), environmental valuation (5\%), and other fields (7\%). In terms of self-assessed expertise, 40\% rated themselves at a medium level, 25\% at medium-high, with smaller percentages at low levels. Scholarly impact scores ranged from 0.0 to 20.0, with 8.0 being the most common score (20\%). The gender distribution was predominantly male (68\%), with 32\% female or undisclosed. Finally, participants reported their primary modelling approaches, MNL and MMNL models were the most commonly used. These were followed by LC models, while other approaches, such as Integrated Choice and Latent Variable, Multiple Discrete-Continuous Extreme Values, and Ordered Logit models, were reported less frequently.

\section{Data analysis}\label{sec:Data_analysis}
In this section, we focus on analysing the choice modellers’ decision-making behaviour captured through the DCM-SG, utilising a behavioural observation framework
\citep{bakeman1997observing}. This approach allows us not only to discuss common methods and analyses employed by participants, but also to identify sequential patterns in their model specification and refining process. We then clustered participants to explore how different factors of in-game usage, the transition between research phases and modelling patterns affect their reported modelling results.

\subsection{Workflows and time spent}\label{sec:5.1}
\begin{figure}[h!]
    \centering
    \includegraphics[width=\textwidth]{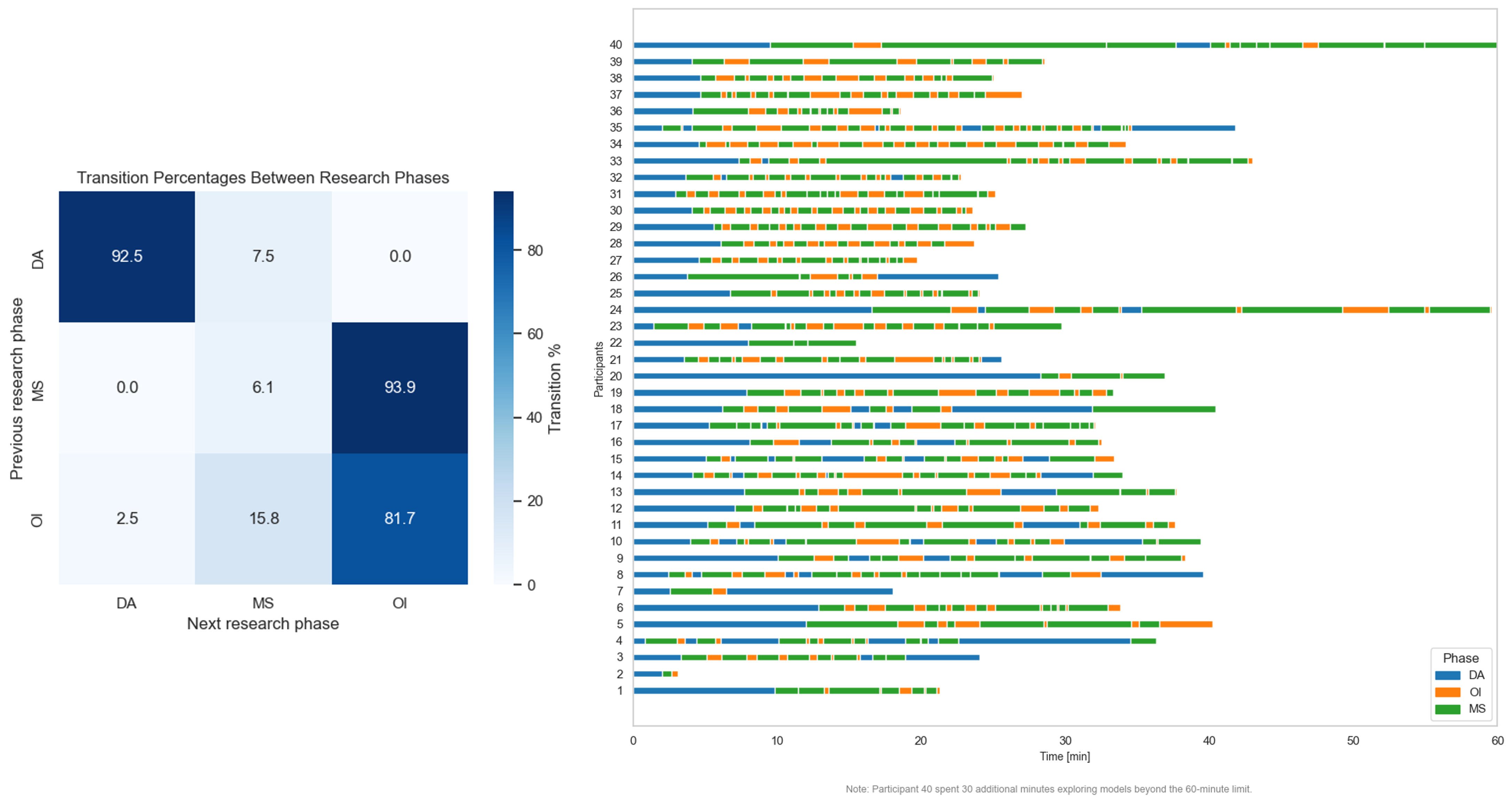}
    \caption{Workflow transitions and time allocation in choice modelling phases}
    \label{fig:workflows}
\end{figure}

We begin our analysis by examining the choice modellers' workflow, focusing on the transitions between research phases and the time allocated to each, as shown in Figure \ref{fig:workflows}. The high intra-phase retention rates, such as 93\% in DA, and 82\% in OI, indicate that modellers typically use all available in-game tools within a phase before gradually moving on to the model specification. Also, the observed intra-phase transitions within the MS phase (6\%) suggest that modellers decide to specify a new model without considering other modelling results than the estimated parameters. This behaviour may reflect a perceived misspecification or unexpected estimation results that lead to a revision of the specification. Notably, few modellers (only 2.5\% of the transitions) reanalyse the data after moving on to model specification and outcome interpretation, which suggests an iterative process concentrated mainly on these two phases. \\

\noindent On the right-hand side, we display the timeline of each choice modeller, representing their progress through the research phases. The graph shows that DA tends to dominate the early workflows, with modellers typically  exploring, visualising, and cleaning data before moving on to more complex phases. A significant amount of time is then spent on MS, reflecting efforts to refine models and incorporate assumptions based on DA and OI analysis. Interestingly, there are instances where modellers briefly return to earlier phases, as discussed before. This highlights the iterative nature of the process, where modellers may revisit the DA phase to re-adjust model hypotheses or perform further descriptive analysis based on new insights from subsequent phases.\\

\subsection{In-game tools analysis}\label{sec:5.1}

Table \ref{tab:combined_in_game} provides a summary of the in-game tools used by participants throughout all research phases. This table shows the number of users who used each tool at least once, the average frequency of use, and the interaction percentage, which reflects the proportion of use within each specific research phase.\\

\noindent In the descriptive analysis phase, participants demonstrated a strong preference for data visualisation tools and statistical descriptions, which can facilitate a deeper understanding of the data prior to model specification. In particular, we observed a significant use of the histograms (19.17\% of interactions in this phase), followed by data dictionary visualisation (13.27\%), box plots and pie charts (both 7.11\%), and correlation matrix plot (6.76\%). Although most participants viewed the main data statistics and the percentage of choice between alternatives to evaluate the data distribution among choice-makers, only approximately 73\% of the modellers (29 out of 40) deleted missing values, and 33\% (13 out of 40) replaced them  before moving to model specification. Some participants chose to explore both options, which were allowed within the game, as part of their data preparation process. Thus, while the use of visualisation tools can facilitate obtaining valuable information for model specification, the exact methodology used to handle the database prior to modelling, even in cases where there is a small amount of missing values, remains unclear. \\

\begin{table}[h!]
\centering
\caption{Summary of in-game tools used across research phases}
\label{tab:combined_in_game}
\begin{tabular}{lccc}
\hline
\textbf{Tools} & \textbf{Users} & \textbf{Mean (SD)} & \textbf{Interaction [\%]} \\ \hline
View summary statistics         & 38 & 1.84 (0.97) & 6.07   \\ 
View data dictionary            & 40 & 3.92 (3.12) & 13.27 \\ 
Check missing data              & 39 & 1.56 (0.75) & 5.29   \\ 
View first 5 rows of data       & 37 & 2.00 (2.39) & 6.42    \\ 
View percentage of choices      & 36 & 1.33 (0.89) & 4.16    \\ 
View choice task example        & 36 & 1.47 (1.00) & 4.60   \\ 
View histogram                  & 34 & 6.50 (4.98) & 19.17  \\ 
Delete missing values           & 29 & 1.21 (0.41) & 3.04    \\ 
View boxplot                    & 26 & 3.15 (2.17) & 7.11   \\ 
Sort dataset by variable        & 22 & 2.14 (1.46) & 4.08   \\ 
View correlation                & 25 & 3.12 (1.64) & 6.76   \\ 
View two-variables scatter plot & 17 & 3.24 (1.92) & 4.77   \\ 
Replace missing values          & 13 & 1.23 (0.44) & 1.39   \\ 
View pie chart                  & 13 & 6.31 (7.90) & 7.11    \\ 
View bar chart                  & 19 & 3.68 (3.00) & 6.07 \\ \hline
Multinomial logit model         & 40    & 5.92 (3.04)  & 51.68   \\ 
Latent class model (2 classes)  & 31    & 2.45 (1.48)  & 17.00   \\ 
Latent class model (3 classes)  & 26    & 1.73 (1.08)  & 10.07  \\ 
Panel mixed logit model         & 27    & 2.26 (1.51)  & 13.64   \\ 
Model misspecification          & 16    & 2.12 (1.85)  & 7.61   \\ \hline
View final log-likelihood       & 33	& 6.58	(4.24)	& 15.16  \\ 
View initial log-likelihood     & 27    & 2.44 (1.72)   & 4.61  \\ 
Calculate Willingness-to-Pay    & 33	& 6.58	(4.24)	& 15.16  \\ 
Model comparison                & 33    & 6.21 (4.54)  & 14.33  \\ 
View number of parameters       & 24    & 3.12 (3.05)  & 5.24   \\ 
View number of individuals      & 17    & 2.00 (1.41)  & 2.38   \\ 
View log-likelihood at equal shares & 21    & 2.43 (1.83)  & 3.56   \\ 
View $\rho^2$                   & 23    & 5.22 (4.48)  & 8.39   \\
View adj. $\rho^2$              & 25    & 5.67 (4.86)  & 10.06    \\
View number of CPU cores        & 17    & 2.18 (1.55)  & 2.59   \\ 
View number of data rows        & 17    & 2.00 (1.41)  & 2.38   \\ 
View number of outputs          & 15     & 2.27 (1.53)  & 2.38   \\ 
View Akaike Information Criterion & 20     & 6.25 (5.09)  & 8.74   \\ 
View Bayesian Information Criterion & 20     & 6.20 (5.07)  & 8.67   \\ 
View time taken for estimation    & 10     & 1.20 (0.42)  & 0.84   \\ \hline
\end{tabular}
\end{table}

\begin{figure}[h!]
    \centering
    \includegraphics[width=0.8\textwidth]{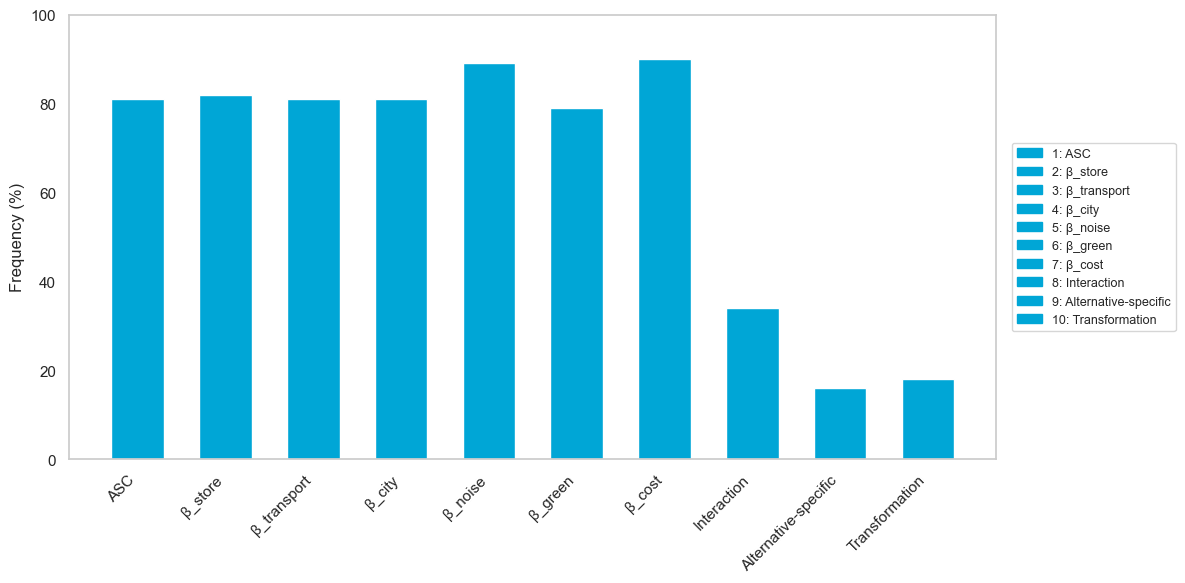}
    \caption{Multinomial logit model specifications among participants}
    \label{fig:mnl}
\end{figure}

\noindent In the model specification phase, participants attempt to specify the three model families available in this SG: MNL, MMNL, and LC models. While many of these were successfully estimated, some specification failed to converge due to misspecification (such as incorrect functional forms, including sociodemographic variables do not vary across utilities, etc.) or because the raw choice dataset lacked sufficient variability to identify certain model parameter. As shown in Table \ref{tab:combined_in_game}, we subsequently analysed each model family to identify the most common specifications.

\noindent Figure \ref{fig:mnl} highlights trends within the specification of MNL models. Despite working with a stated preference database with unlabelled alternatives, approximately 80\% of the specified models included alternative-specific constants (ASCs) and most did not incorporated all attributes. This suggests that choice modellers were attentive to possible differences in baseline utilities. As their estimated model results show, ASCs were often statistically significant and improved model performance, which possibly reflects efforts to account for design effects or lexicographic behaviours in the decision-makers’ choice process. In general, attribute effects were included as linear-additive, the Store attribute in 82\% of the models, Transport in 82\%, City in 82\%, Noise in 88\%, Green in 79\%, and Cost in 89\%. Transformations of attributes were limited, with only 18\% of the specifications including logarithmic or Box-Cox transformations. Furthermore, while the majority of attributes were treated as generic, which is consistent with the context of unlabelled alternatives, approximately 16\% of the cases were tested with alternative-specific taste parameters.  Regarding model interactions, 34\% of the specified models include at least one attribute-covariate interaction. The most frequently observed were between Transport, City, Noise, and Green with the respondent's Age, as well as interactions between Store and gender (Woman), and between Cost and homeonwership status (Homeowner). However, there was a limited exploration of interactions related to the residence location (respcity), even though the data were collected in four different cities. This evidence reveals a clear preference for linear parameter specifications, with a tendency to test for attribute exclusion and  less frequently, the integration of sociodemographic characteristics. This pattern may reflect limitations in the model specification phase of the game, which did not allow for modelling categorical attributes as dummy variables. This limitation may have prevented modellers from exploring other ways they could have considered. Furthermore, the limited exploration of interactions may be due to time constraints, which prevented modellers from fully investigating hypotheses related to preference variations of decision-makers across cities.\\

\begin{figure}[h!]
    \centering
    \includegraphics[width=\textwidth]{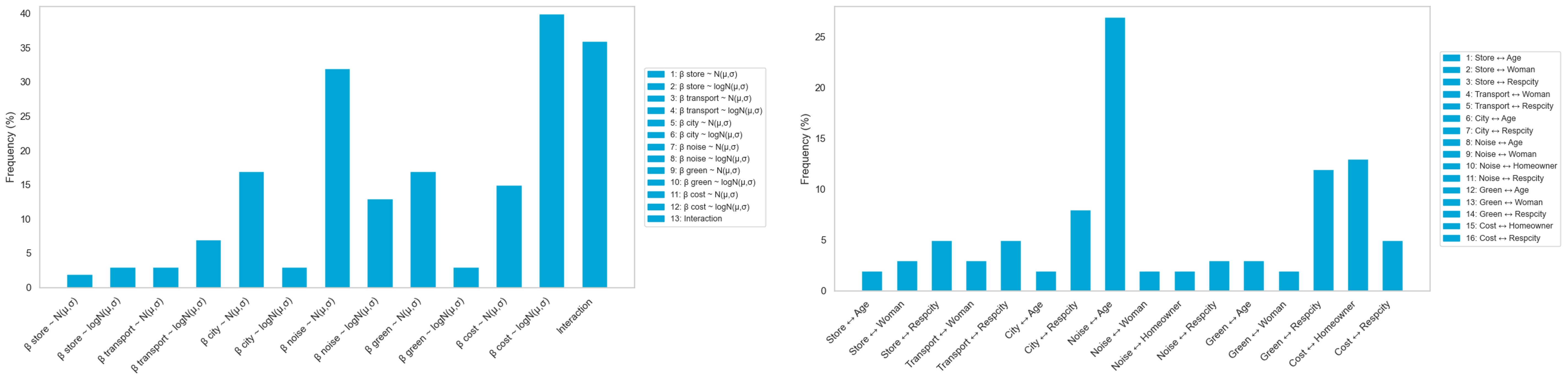}
    \caption{Mixed multinomial logit model specifications among participants}
    \label{fig:mmnl}
\end{figure}

\noindent Figure \ref{fig:mmnl} shows the MMNL model specifications, wherein all attributes were considered linear and generic by default. This restriction imposed by the DCM-SG does not seem to have affected this analysis, as it is consistent with the specifications observed in the MNL models. In terms of parameter distribution, 44.3\% of specifications considered a single distributed parameter, with 55.5\% of them following a normal distribution and the remainder log-normally distributed. When two random parameters were included, 44.9\% were modelled as normally distributed, 35.6\% as log-normally distributed, and 19.5\% as a combination of both distributions. Notably, Noise and Cost were the most frequently specified attributes with normal and log-normal distributions, respectively. Moreover, 63.9\% of the models did not include interactions with sociodemographic variables, with city being the least frequently considered in such interactions. When interactions were included, the most common were with age, followed by residence location and gender, though their frequency remained low. These findings evidence an approach focused on capturing taste variations for noise and cost along with an exploration of the observed heterogeneity. However, considering a normal distribution for the cost attribute has important implications, since it always results in undefined Willingness-to-Pay values for noise, a problem that can be corrected by using log-normal distributions, which restrict the values to a nonnegative space and provide greater consistency in the economic interpretation of the results.\\

\begin{figure}[h!]
    \centering
    \includegraphics[width=\textwidth]{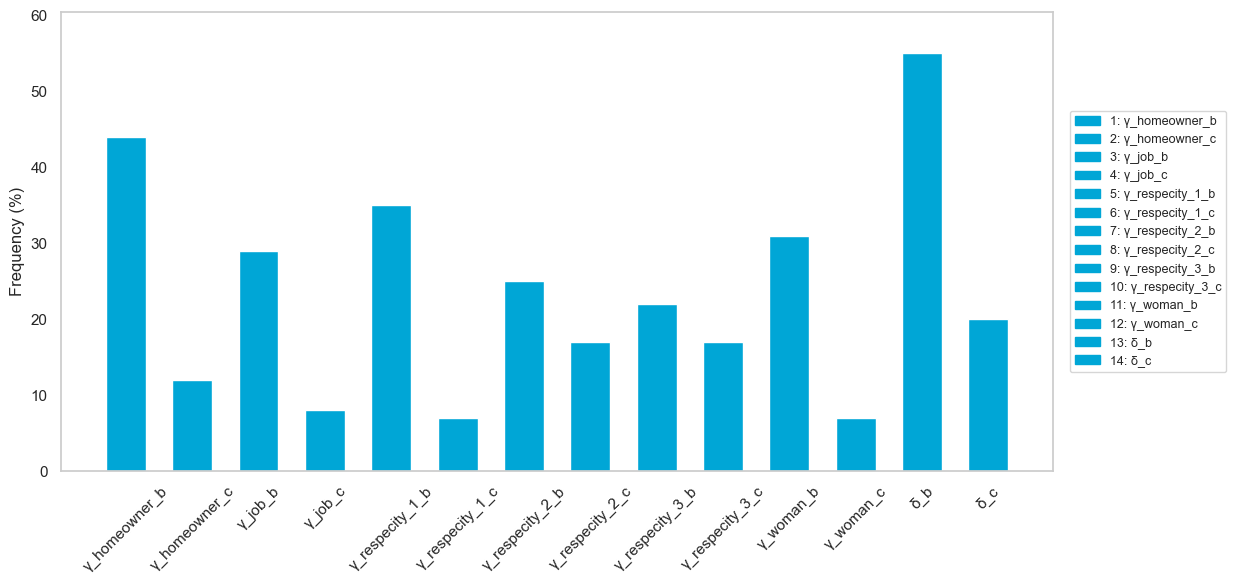}
    \caption{Latent class model specifications among participants}
    \label{fig:lc}
\end{figure}

\noindent Figure \ref{fig:lc} shows the specifications of the LC model, where, as in the previous case, all attributes were considered linear for each class by default. In the modellers’ decisions, 55.3\% of the specifications involved latent class models with two classes, while the remaining models utilised three classes. This suggests that many modellers preferred to work with less complex models that may be easier to interpret and estimate. Moreover, only six modellers were able to successfully specify three-class latent models despite several trials, indicating that greater complexity is involved in specifying LCs with real-world data. Also, due to the early issues of parameter non-identification, modellers tended to focus on models with fewer classes or other model families. In terms of the membership function, the inclusion of covariates such as Homeowner, Woman, Job, and Respcity varies between 8\% and 44\%, which demonstrates an attempt to capture socio-demographic heterogeneity in the membership class function.  However, the relatively low usage rates of Respcity in three-class models indicate that modellers may not have fully explored the potential influence of location-based heterogeneity. This aligns with observations in the MNL and MMNL models, where interactions with residential locations were also limited.\\

\noindent In the outcome interpretation phase, when modellers estimated a model, they were immediately given a standard result table generated by the DCM-SG, which included the parameter names, estimated values, robust standard errors, t-tests against zero, and corresponding p-values. Analysis of the tools for interpreting outcomes revealed that modellers focused mainly on the calculation of Willingness-to-Pay (15.16\% of the interactions) and on comparison between models (14.33\% of the total interactions). Similarly, the observation of log-likelihood (13.07\%) was also a highly reviewed, which indicates the importance modellers' attach to the internal consistency of the model and its predictive capability. However, modellers paid limited attention to other metrics such as the (adjusted-) rho square, the Akaike Information Criterion, and the Bayesian Information Criterion, each used in approximately 9\% of interactions. These preferences not only suggest that model refinement was guided by goodness-of-fit and economic interpretation of the parameters, using the previous model as a benchmark for comparison, while neglecting metrics that assess the trade-off between model fit and model complexity, which is essential for selecting robust and parsimonious models.\\

\begin{figure}[h!]
    \centering
    \includegraphics[width=\textwidth]{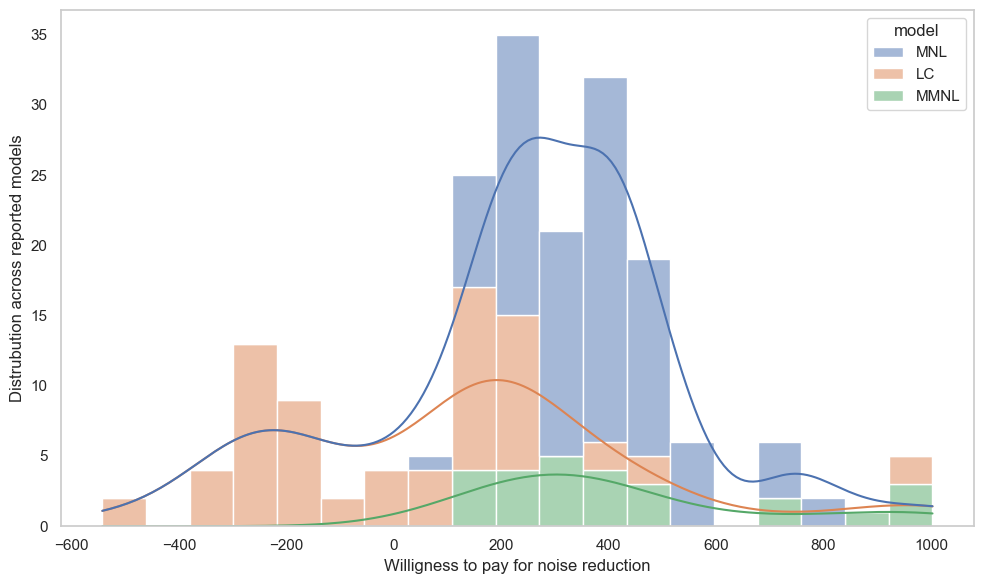}
    \caption{Workflow transitions and time allocation in choice modelling phases}
    \label{fig:wtp}
\end{figure}

\noindent Finally, in the reporting phase, Figure \ref{fig:wtp} shows the Willingness-to-Pay reported by modellers towards the conclusion of the simulated research, which reveals important implications for informing policymakers. On the one hand, MNL and MMNL models show consistent positive WTP estimates for noise reduction, indicating a generalised preference among case study decision-makers. Although latent class models (LC2 and LC3) uncover significant preference variations, with some classes showing negative WTP values, MNL was the most common type of model reported (55\% of final reports), followed by LC models (32.5\%) and finally MMNL (12.5\%). This heterogeneity in the results not only highlights the variety of results that modellers found acceptable, but also reflects a complex decision-making process in their final reports. Also, this clear distinction in reported WTP values between models serves as a reminder to choice modellers of the importance of considering multiple modelling approaches to fully capture observed and unobserved heterogeneity in preferences.

\subsection{Analysis of model specification workflows}\label{sec:model_spec_workflows}

\begin{figure}[h!]
    \centering
    \includegraphics[width=\textwidth]{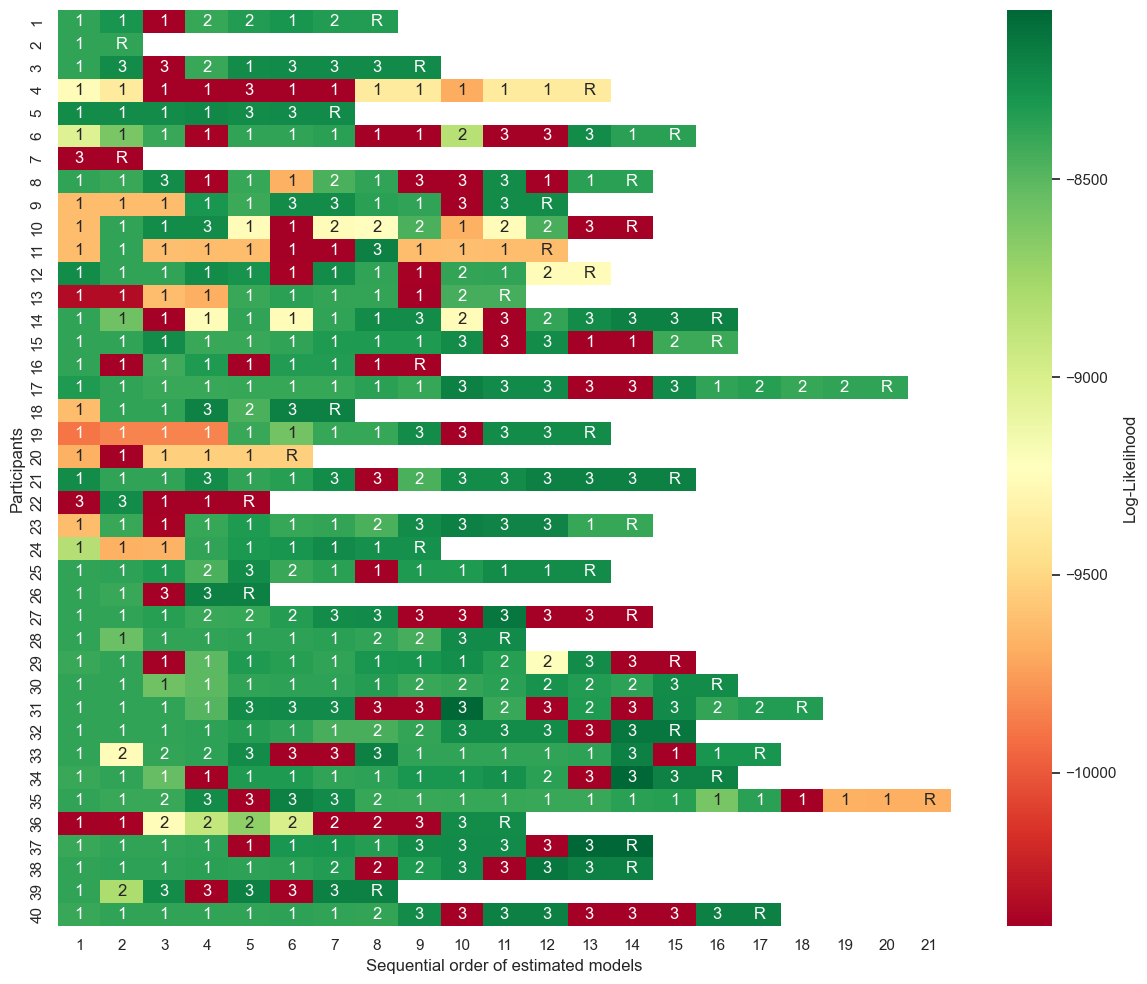}
    \caption{Workflow transitions and time allocation in choice modelling phases}
    \label{fig:ms_wk}
\end{figure}

To gain deeper insights into the modelling workflows employed by the participants during the model specification phase, we display the temporal progression of models in the Figure \ref{fig:ms_wk}). This graph provides an aggregated view of how modellers transitioned across different model families, such as MNL, LC, and MMNL, as well as cases of Misspecification (Miss), and the point at which they reported their final model (R). The diagram shows the evolution of the estimated models and their Log-likelihood values, starting from their initial specification to the reported one.\\ 

\noindent As can be seen, during the initial stage of the modelling process, most participants (38 out of 40) began by specifying MNL models, which varied in terms of the attributes included, the transformations applied, and the interactions specified with sociodemographic variables. Notably, only four participants started with a fully linear specification, including all attributes. This may seem a counterintuitive phenomenon, as one might expect modellers to begin by including all available variables from the stated choice tasks to capture primary and linear effects on choices. This behaviour may reflect the influence of preconceptions about data complexity or prior experiences with certain model types. As participants progressed, more complex functional forms were tested within the MNL family, some of which resulted in misspecifications. This may have been due to attempts to include more intricate relationships with sociodemographic variables and transformations to attributes to explain observed heterogeneity, leading to the non-identification of parameters.\\

\noindent In later iterations, there was a shift in the trend towards more complex models, with participants slowly specifying LC and MMNL models. This evolution suggests a learning trajectory, where initial explorations with simpler MNL models laid the groundwork for the adoption of more flexible models capable of accounting for deterministic and random taste heterogeneity. This pattern aligns with standard modelling practice, where simpler models are frequently estimated as a benchmark before progressing to more advanced approaches.\\

\subsection{Workflows analysis difference}\label{sec:5.3}

To evaluate the impact of participants’ workflows on the improvement of choice modelling outcomes, we conducted an analysis of modelling patterns. Using the cSPADE algorithm, a data mining method designed to identify frequent patterns within temporal sequences \citep{zaki2001spade, maimon2005data}, we examined the most common patterns related to in-game tool usage, transitions between research phases, and shifts between different model families.\\

\noindent Firstly, participants were classified into two groups based on changes in goodness-of-fit metrics, such as Log-likelihood, AIC, BIC, $\rho^2$, and ${\bar{\rho}}^2$, from the initial model to the final reported model. The first group included 30 participants who achieved improvements in these metrics, suggesting  a target learning process to progressively capture variability in the choices through more complex model specifications. The second group consisted of 10 participants whose reported models did not surpass the initial metrics, which may indicate challenges in refining their modelling approaches or limitations in their ability to adapt their modelling approach to better fit the data. Secondly, the most frequent patterns of in-game tool usage, transitions between research phases, and shifts between model types were found considering the cSPADE algorithm with minimum support thresholds of 70\%. Thus, only sequential patterns that appeared in at least 70\% of the observed workflows were included in the analysis. This value was chosen to reduce noisy patterns and extract the most common behavioural workflows within our small-to-medium sample, as suggested by  \citep{zhang2023sequential, kang2017using}. Finally, we calculated the frequency of each pattern for every choice modeller and compared the two groups using an independent samples t-test.\\

\begin{table}[h!]
\centering
\caption{Difference across frequent patterns (p-value $<$ 0.05)}
\label{tab:pattern_t_tests}
\begin{tabular}{lcc}
\hline
\textbf{In-game tools and workflows} & \textbf{t-statistic} & \textbf{p-value} \\ \hline
 Total uses: View summary statistics & -2.82 & 0.055 \\ 
 Total uses: View bar plot & -2.20 & 0.042 \\ 
 Total uses: View correlations & -2.69 & 0.013 \\
 Total uses: Model comparison with previous one & -2.20 & 0.046 \\
 Total uses: View Akaike Information Criterion & -5.38 & 0.001 \\ 
 Total uses: Handling missing data & -3.21 & 0.002 \\ 
 Total uses: Replacing missing values & 2.94 & 0.011\\
 
 Time spent on visualising histograms & -2.82 & 0.027 \\
 Time spent on specifying MNL & -2.64 & 0.013 \\ \hline
 Calculate WTP $\rightarrow$ MNL & -2.75 & 0.009 \\
$\bar{\rho}^2 \rightarrow$ Calculate WTP & -2.73 & 0.011 \\
MNL $\rightarrow$ Log-Likelihood & -3.14 & 0.003 \\
Calculate WTP  $\rightarrow$ Model comparison with previous one & -3.03 & 0.005 \\
BIC $\rightarrow$Calculate WTP & -2.81 & 0.009 \\ \hline
Miss $\rightarrow$ Miss & -3.10 & 0.004 \\
MMNL $\rightarrow$ LC & -2.69 & 0.013 \\ \hline
DA $\rightarrow$DA $\rightarrow$ MS $\rightarrow$ OI $\rightarrow$ OI $\rightarrow$ MS & -3.53 & 0.001 \\
MS $\rightarrow$ OI $\rightarrow$ OI $\rightarrow$ OI $\rightarrow$ OI & -2.96 & 0.006 \\
OI $\rightarrow$ OI $\rightarrow$ OI $\rightarrow$ OI $\rightarrow$ MS & -3.18 & 0.003 \\
OI $\rightarrow$ OI $\rightarrow$ DA $\rightarrow$ DA $\rightarrow$ MS & -2.26 & 0.031 \\ \hline
\end{tabular}
\end{table}

\noindent Table \ref{tab:pattern_t_tests} shows statistical differences between the two groups of participants in terms of the use of in-game tools, transitions between research phases, and model specification workflow patterns. Participants who used  tools such as ‘View summary statistics’,  ‘View bar plot’ and ‘View correlations’, or who spent more time in ‘Visualising histograms’ or ‘Specifying MNL models’, were more likely to report improved modelling metrics compared to their initial estimations.  Moreover, specific workflow sequences such as moving from ‘Handling missing values’ to ‘Viewing summary statistics’ to ‘viewing the first five rows’, as well as revisiting the Descriptive Analysis phase after interpreting their modelling outcomes (OI → DA), were more common among modellers who refined their specifications over time. These patterns of behaviour suggest a higher involvement in thorough data analysis and development of initial hypotheses, which therefore contributed to improved models.\\

\noindent  Another important distinction emerged in how participants analysed modelling outcomes. Those who adopted a comprehensive approach, considering not only parameter estimate values and their standard deviations, but also model fit metrics such as $\rho^2$, $\bar{\rho}^2$, AIC, and BIC, were more likely to improve their modelling outcomes. In addition, the workflow `Calculate WTP → Model comparison with previous one’ reflects a deliberate focus on both economic interpretation and comparative model evaluation, often leading to better model refinement. Similarly, participants who engaged in sequences such as ‘AIC → Calculate WTP’ or ‘BIC → Calculate WTP’ demonstrated significantly better model improvements than those who did not. These patterns suggest that integrating statistical diagnostics with economic reasoning supported more informed and effective model development.\\

\noindent In terms of transitions between modelling phases, significant differences in workflows appeared when several DA and OI phase tools were used before specifying a model. For instance, sequences such as `DA $\rightarrow$ DA $\rightarrow$ MS $\rightarrow$ OI $\rightarrow$ OI $\rightarrow$ MS' and `DA $\rightarrow$ MS $\rightarrow$ OI $\rightarrow$ OI $\rightarrow$ MS' suggest that revisiting  data analysis before and after initial model specification further improved model refinement. Similarly, extended use of OI tools prior to respecification, such as `MS $\rightarrow$ OI $\rightarrow$ OI $\rightarrow$ OI $\rightarrow$ OI'  and `OI $\rightarrow$ OI $\rightarrow$ OI $\rightarrow$ OI $\rightarrow$ MS',  was associated with improved model fit, whereas those who relied solely on estimation tables showed minimal or no improvement. Moreover, the pattern `OI $\rightarrow$ OI $\rightarrow$ DA $\rightarrow$ DA $\rightarrow$ MS' further supports that idea of returning to data exploration after interpreting results can inform initial modelling hypotheses, and ultimately, lead to improved model performance. These results suggest that multiple specifications followed without proper analysis of the modelling results may lead to a search for models that ultimately do not improve upon the initial model.\\ 

\noindent Finally, we found that participants who achieved improvements in model fit also tended to exhibit a higher frequency of model misspecification transitions (Miss → Miss) and also to transitions from MMNL to LC models. This behaviour suggests a continued effort to capture more complex functional forms and reflects an iterative learning process throughout the modelling process. As shown in Figure \ref{fig:ms_wk}, the workflows of participants 4, 17, and 36, among others, demonstrate that repeated misspecifications were often followed by model specifications that achieved better performance within the same model family.\\ 

\noindent Overall, the results indicate that participants who engaged in thorough data analysis, revisited earlier research phases to refine their modelling assumptions, and evaluated goodness-of-fit metrics were more successful in reporting models with an improved balance between fit and parsimony. These findings tangibly reflect the nature of discrete choice modelling as an intrinsically iterative, hypothesis-driven, and feedback-guided process, which relies not only on model fit metrics, but also on alignment with expected behavioural realism. 

\section{Conclusions}\label{sec:conclusions}

Our study provides a twofold contribution to the choice modelling field. First, it introduces serious games as a methodological innovation to capture workflows of choice modellers and demonstrates how serious game data can effectively be analysed to better understand their decision-making process. The Discrete Choice Modelling Serious Game provides an online environment that simulates actual research phases, enabling modellers to apply their knowledge while we track their actions. We have made our code openly available to facilitate future research using this tool. \\

\noindent Second, our study provides new substantive insights into the practices of choice modellers. On the one hand, we found strong evidence of the iterative nature of the choice modelling process, with choice modellers moving back and forth between the modelling phases such as descriptive analysis, model specification, outcome interpretation and reporting phase. For instance, we have observed that - after the first descriptive analysis - most participants start with specifying MNL models, after which they either return to the descriptive analysis phase or move forward to more advanced models, such as MMNL and LC. Furthermore, we have found extensive support for the notion that modelling practices are heterogeneous. Specifically, we found that while data visualisation and statistical descriptions were commonly used, there was no clear approach to handling missing values; We also observed participants favoured simpler models despite having complex families available in the game. On the other hand, our results also reveal that workflows, in-game tools usage, and model specification strategies significantly impact choice modelling outcomes. Participants who engaged in comprehensive data exploration, and iterative comparisons, and made systematic use of econometric tools tended to improve goodness-of-fit and parsimony. Conversely, those who relied more heavily on limited metrics without exploring broader aspects of the model struggled to improve upon their initial specifications. \\

\noindent Despite these contributions,  our study has several limitations that future research should address. The current structure of the serious game, specifically in the model specification phase, imposes constraints that do not totally replicate real-world modelling scenarios. While the game is designed to support iterative exploration, the set of tools and model options remains limited compared to what modellers may use in practice. This may influence participants’ behaviour, strategies,  and decision-making processes they might otherwise pursue. Additionally, the preference for simpler models observed in this analysis (63\% of participants selected MNL model as their reported model) may be partially influenced by the time constraints assigned to the game session. In the real world, modellers typically have more time and flexibility to explore complex model families and refine specifications. Moreover, the design of the game may have encouraged participants to prioritise completing the task over engaging in extended model exploration. Furthermore, although the small-medium sample size limits the generalisability of our results, this study serves as a methodological starting point and demonstrates the potential of the DCM-SG to capture meaningful modelling behaviours.  Finally, participants’ awareness that they were being monitored may also have introduced a bias, leading them to display desirable behaviours rather than their own modelling practices. For instance, while we observed some modellers revisiting earlier phases (e.g., returning to DA), it remains unclear whether such behaviour mirrors real-world practice or was influenced by the experimental setup.\\

\noindent To overcome these limitations, future research should involve larger samples with more diverse backgrounds, extend the duration of game sessions to allow deeper engagement with advanced models, and improve in-game tools to enable more complex approaches in the model specification phase.

\section{Acknowledgements}
We would like to express our sincerest gratitude to all participants of our serious games. Their willingness, dedication, and active involvement in the game are greatly appreciated. In particular, we acknowledge the following individuals for their significant contributions and valuable insights, which greatly enhanced the research process: Bartosz Bursa, Victor Cantillo, Santiago Cardona, Laurent Cazor, Roel Faber, Sayed Faruque, Francisco Garrido-Valenzuela, Thomas Hancock, Bastian Henriquez-Jara, Chengxi Liu, Gregory Macfarlane, Evripidis Magkos, Petr Mariel, David Palma, Pablo Reyes, Jaime Soza-Parra, Melvin Wong, Amir Ghorbani, Baiba Pudāne, Peter King, Edward JD Webb, Nejc Geržinič, Rodrigo Tapia, and Marco Kouwenhoven. Stephane Hess acknowledges support from the European Research Council through advanced grant 101020940-SYNERGY. 

\clearpage
\section*{Appendix}


\clearpage

\end{document}